\documentclass[pdflatex,sn-nature]{sn-jnl}

\usepackage{multirow}
\usepackage{amsmath,amssymb,amsfonts}
\usepackage{amsthm}
\usepackage{mathrsfs}
\usepackage[title]{appendix}
\usepackage{xcolor}
\usepackage{textcomp}
\usepackage{manyfoot}
\usepackage{booktabs}
\usepackage{algorithm}
\usepackage{algorithmicx}
\usepackage{algpseudocode}
\usepackage{listings}
\usepackage{graphicx}
\usepackage{dcolumn}
\usepackage{bm}
\usepackage{xcolor}
\usepackage{setspace}
\usepackage{textcomp, gensymb}
\usepackage{soul}
\usepackage{fancyhdr}
\usepackage{booktabs}
\usepackage{upgreek}
\usepackage[version=4]{mhchem}
\usepackage{gensymb}
\usepackage{lineno}
\begin{document}
\title[Article Title]{Magnetotransport signatures of spin-orbit coupling in high-temperature cuprate superconductors}

\author[1]{\fnm{Aleix} \sur{Barrera}}

\author[1]{\fnm{Huidong} \sur{Li}}

\author[1]{\fnm{Thomas} \sur{Gunkel}}

\author[1]{\fnm{Jordi} \sur{Alcalà}}

\author[1]{\fnm{Silvia} \sur{Damerio}}

\author*[1]{\fnm{Can Onur} \sur{Avci}}\email{cavci@icmab.es}

\author*[1]{\fnm{Anna} \sur{Palau}}\email{apalau@icmab.es}

\affil[1]{\orgdiv{Institut de Ciència de Materials de Barcelona, ICMAB-CSIC},\orgaddress{\street{Campus UAB}, \city{Bellaterra, Barcelona}, \postcode{08193}, \country{Spain}}}

\abstract{\textbf{Spin transport in superconductors offers a compelling platform to merge the dissipationless nature of superconductivity with the functional promise of spin-based electronics. A significant challenge in achieving spin polarisation in conventional superconductors stems from the singlet state of Cooper pairs, which exhibit no net spin. The generation of spin-polarised carriers, quasiparticles, or triplet pairs in superconductors has predominantly been realised in hybrid superconductor/ferromagnet systems through proximity-induced spin polarisation. Historically, cuprate superconductors have been characterised by strong electronic correlations but negligible spin-orbit coupling. Here, we report exceptionally large anisotropic magnetoresistance and a pronounced planar Hall effect arising near the superconducting phase transition in the prototypical high-temperature cuprate superconductor \ce{YBa2Cu3O_{7-x}} without using a proximity ferromagnet. These effects —unprecedented in centrosymmetric cuprates— emerge from spin-polarised quasiparticle transport mediated by strong spin-orbit coupling. By systematically tuning magnetic field strength, orientation, temperature, and doping, we show clear evidence of spin-orbit-driven transport phenomena in a material class long thought to lack such interactions. Our findings reveal an unexpected spin-orbit landscape in cuprates and open a route to engineer spintronic functionalities in high-temperature superconductors.}}


\maketitle

Unusual magnetotransport phenomena at phase transitions provide a powerful lens for exploring the intricate interplay between charge, spin, and lattice degrees of freedom in condensed matter systems. Classic examples include colossal magnetoresistance in correlated oxides \cite{Ramirez_CMR1997}, magnetocaloric effects in rare-earth alloys \cite{Phan_Magnetocaloric}, quantum Hall transitions in two-dimensional electron gases \cite{Chakraborty2000}, and topological phase transitions in Weyl semimetals \cite{Yan2017}. More recently, novel spin transport effects have emerged at magnetic phase transitions, such as spin colossal magnetoresistance \cite{Qiu2018} and longitudinal spin pumping \cite{Lee2025}. These phenomena illustrate the dynamic evolution of spin and charge in response to symmetry breaking and the changes in the order parameters, revealing emergent physical effects that govern electronic transport. Unravelling these mechanisms deepens our understanding of fundamental physics and holds promise for next-generation technologies in energy-efficient electronics, quantum materials, and spintronic computing.

Spin transport in superconducting materials represents a particularly intriguing frontier, as it combines the dissipationless nature of superconductivity with the potential advantages of spin-based electronics. This field offers exciting opportunities for energy-efficient information processing, storage, and transmission \cite{YangRobinson2021,Ohnishi2020}. However, generating and manipulating spin polarisation in conventional superconductors is inherently challenging due to the singlet nature of Cooper pairs, which carry zero net spin. Beyond the realm of triplet superconductivity, spin-polarised quasiparticles have emerged as a promising alternative, exhibiting distinct transport properties that differ from those of conventional electrons. In particular, quasiparticle-mediated spin currents could surpass electron-mediated currents in efficiency, owing to their longer spin lifetimes and the possibility of spin-charge separation, enabling nearly dissipationless transport \cite{Kuzmanovic2020,Quay2013,Hubler2012}. Furthermore, quasiparticle-mediated spin Hall effects may induce giant spin-to-charge conversion phenomena \cite{Wakamura2015,Jeon2020}. Notably, in the presence of an in-plane magnetic field, spin-polarised quasiparticles can emerge even from a non-magnetic injector due to Zeeman splitting of the density of states \cite{Bergeret2018,Quay2013,Wolf2013}.

Magnetoresistance and Hall effects provide invaluable tools for probing the electronic properties of complex quantum materials. In superconductors, these techniques serve as sensitive probes of the interplay between competing interactions, often revealing unexpected transport anomalies \cite{Puica2004}. For instance, the Anomalous Hall Effect (AHE) in superconductors may arise from multiple mechanisms, including vortex motion with pinning, superconducting fluctuations\cite{Wordenweber2006,Wang1994}, quasiparticle excitations, or topological scattering effects \cite{Zhao2019,Feigelman1995}. In cuprate superconductors, spontaneous transverse voltages have been observed even at zero applied field, suggesting an intrinsic electronic nematicity \cite{Wu2017}. Additionally, anisotropic magnetoresistance and its transverse counterpart, the Planar Hall Effect (PHE), provide crucial insights into hidden spin symmetries and exotic electronic states driven by spin-orbit coupling (SOC) and topological band structure effects. These transport signatures offer a powerful means to uncover previously hidden physics, particularly in the context of unconventional superconductivity and emergent spin-orbit interactions.

Here, we report exceptionally large anisotropic magnetoresistance and planar Hall effects occurring within a narrow temperature window at the superconducting phase transition in the archetypal high-temperature cuprate superconductor \ce{YBa2Cu3O_{7-x}} (YBCO). Systematic measurements as a function of external magnetic field, temperature, oxygen doping level, and field geometry reveal spin-polarised and quasiparticle-mediated transport governed by strong SOC. These results challenge the long-held assumption that SOC is negligible in centrosymmetric cuprate superconductors. Despite being overlooked until now, recent measurements using spin- and angle-resolved spectroscopy have uncovered interesting signs of SOC effects in bismuth-based superconductors \cite{Gotlieb2018,Iwasawa2023,Luo2024}. Building on these findings, theoretical predictions suggest that various spin-orbit-driven phenomena could occur in cuprate superconductors \cite{Raines2019}. However, direct transport measurements reflecting these effects were missing. Our findings provide new insights into charge and spin transport in high-temperature superconductors and open avenues for developing highly efficient spin-based electronic devices operating at liquid nitrogen temperatures.

\section*{Sample description and planar Hall effect measurements}

\begin{figure}[h]
\centering
\includegraphics[width=0.9\textwidth]{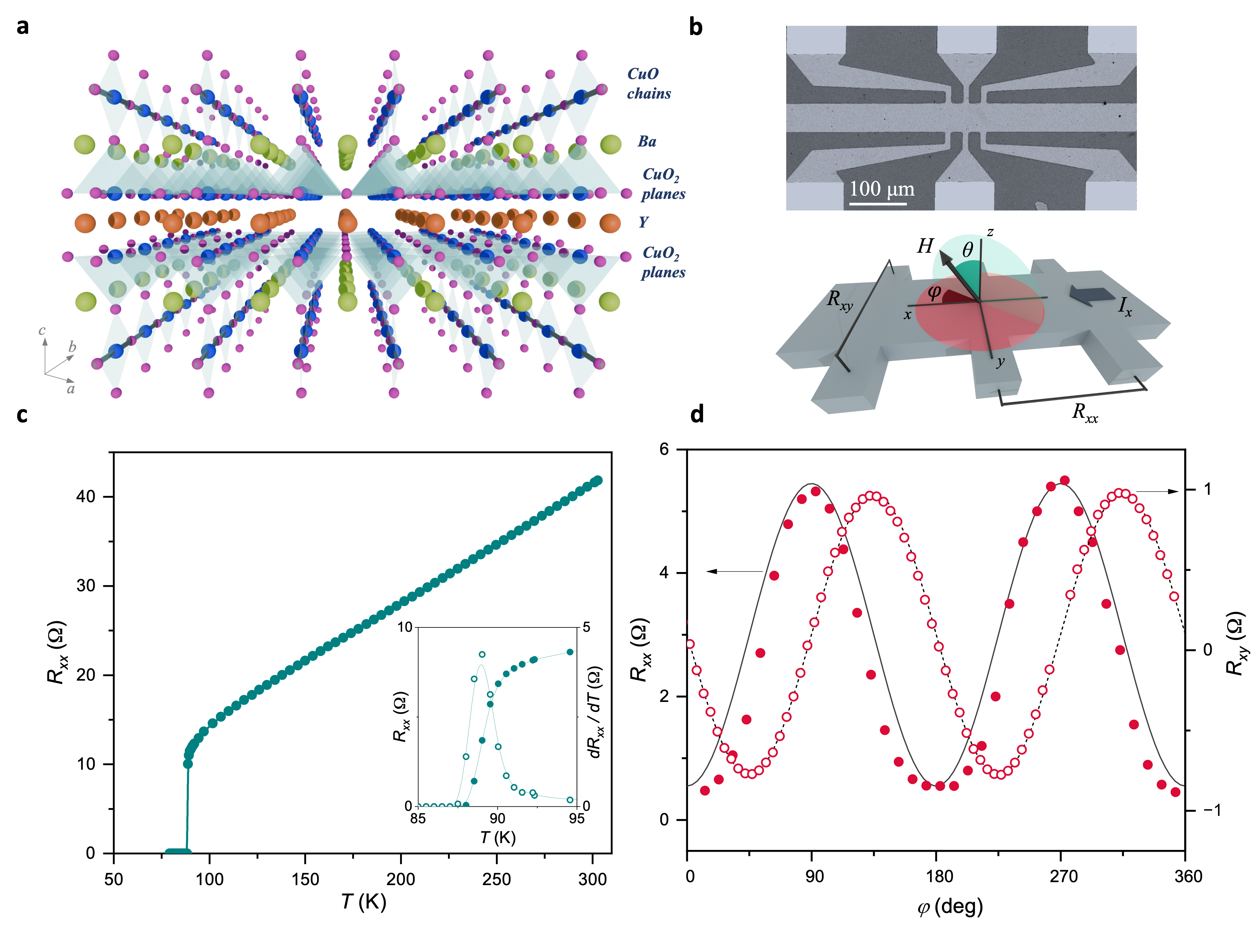}
\caption{\textbf{Layer structure, device layout and magnetotransport measurements.} \textbf{a}, Crystal structure of YBCO. \textbf{b}, Scanning electron micrograph of a sample (top) and schematics of the angular longitudinal and transversal resistance measurements (bottom).  \textbf{c}, Temperature dependence of the longitudinal magnetoresistance measured at zero applied field for a typical YBCO device of 50nm. Inset shows a zoom of the magnetoresistance (closed symbols, left axis) and its first derivative (open symbols, right axis). \textbf{d}, In-plane angular dependence of the magnetoresistance (closed symbols, left axis) and Hall resistance (open symbols, right axis) measured at 88~K and 8~T. Solid and dashed lines show the angular dependence of cos$^2$($\varphi$) and sin($\varphi$)cos($\varphi$), respectively.}
\label{Figure 1}
\end{figure}

We grew epitaxial \ce{YBa2Cu3O_{7-x}} (YBCO) superconducting thin films by pulsed laser deposition on single crystal substrates (details in Methods). YBCO is a layered perovskite high-temperature superconductor with an orthorhombic crystal structure, where superconductivity arises primarily from the \ce{CuO2} bilayers separated by Y and Ba spacer layers. The additional CuO chain layers along the \textit{b}-axis are crucial in charge doping and transport, influencing the anisotropic electronic properties of the material (Figure~\ref{Figure 1}a). The superconducting transition temperature and electronic phase diagram are susceptible to oxygen stoichiometry, with optimal doping \textit{x} = 0.08  yielding $T_c$ = 93 K \cite{Deutscher2014}. The films are patterned in Hall bar structures with a length \textit{l} = 20-30 \micro m, width $\textit{w}$ = 20-50 $\micro$m, as shown in the top panel of Fig.~\ref{Figure 1}b. Hall effect and magnetoresistance measurements described below are performed with electrical connections and field geometries described in the lower panel of Fig.~\ref{Figure 1}b in a Physical Property Measurement Setup. 

Fig.~\ref{Figure 1}c depicts the temperature dependence of the longitudinal resistance ($R_{xx}$) measured in the absence of an external magnetic field for an optimally doped 50 nm-thick YBCO. The superconducting transition is determined at the first derivative's maximum and reads $T_c$ = 89~K, which is slightly below the bulk value and indicates an excellent film quality. We next measure $R_{xx}$ and its transverse counterpart $R_{xy}$ at 88~K, i.e., at the onset of superconductivity, by rotating the magnetic field within the \textit{xy}-plane at an amplitude of 8~T. The results are shown in Fig.~\ref{Figure 1}d and represent the most critical findings of this study. We find an exceptionally large anisotropic magnetoresistance (AMR) in excess of 1000\% and a sizeable planar Hall effect (PHE) resistance -the transverse counterpart of AMR, close to 1 Ohm. Both signals follow the characteristic symmetries typically observed in ferromagnetic materials, i.e., $\propto$ cos$^2$($\varphi$) and $\propto$ sin($\varphi$)cos($\varphi$), respectively, despite the absence of ferromagnetic ordering in YBCO. 

\begin{figure}[h]
\centering
\includegraphics[width=0.9\textwidth]{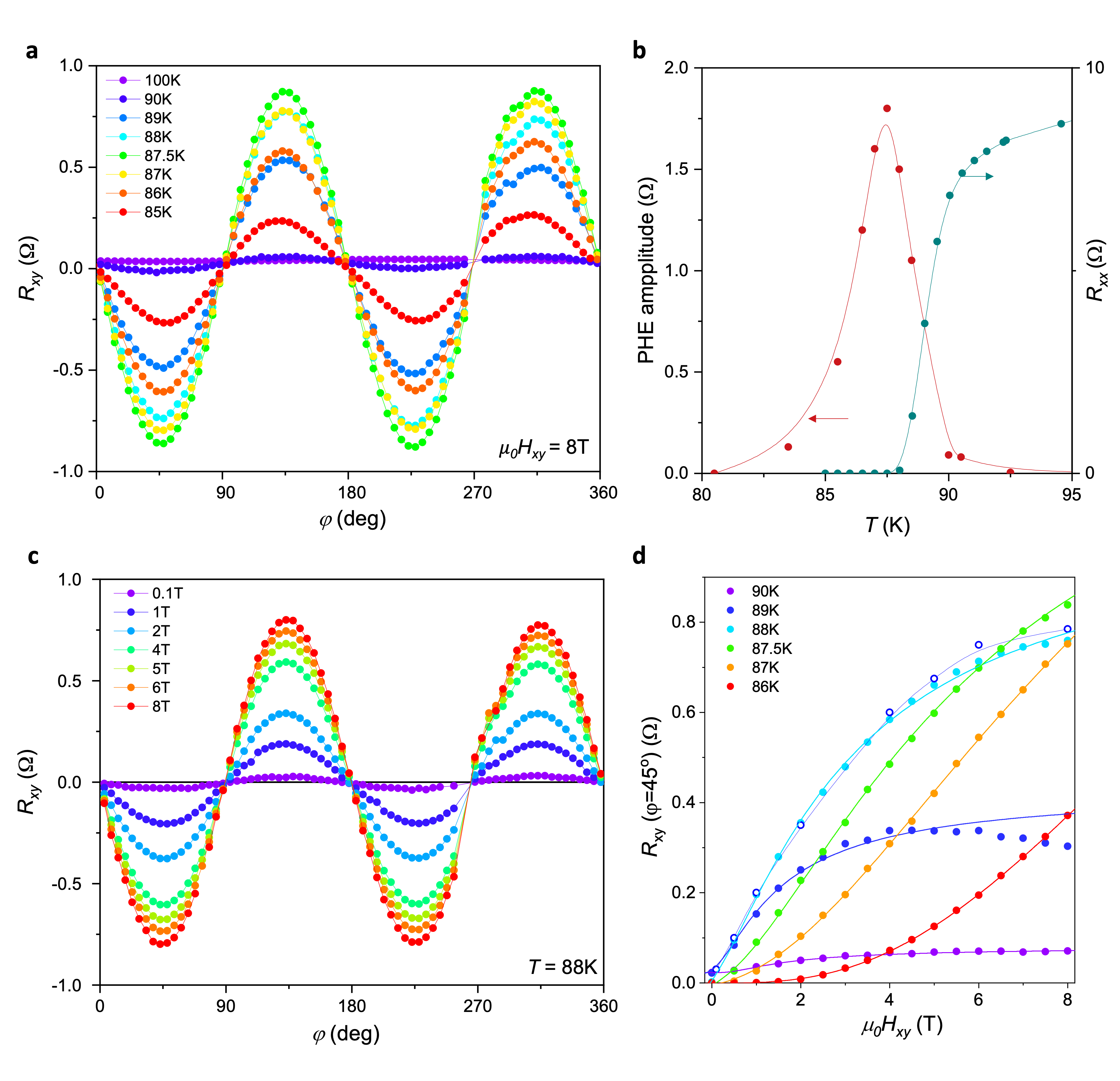}
\caption{\textbf{Angle-dependent Hall resistances}. \textbf{a},\textbf{c}, Angular dependence of the planar Hall resistance, $R_{xy}(\varphi)$ measured at 8~T for different temperatures (\textbf{a}) and 88K for different applied magnetic fields (\textbf{c}) for a typical YBCO device of 50nm. \textbf{b}, Red closed symbols show the temperature dependence of the PHE amplitude obtained by fitting the curves in \textbf{a} with a sin($\varphi$)cos($\varphi$) dependence. Cyan open symbols show the longitudinal magnetoresistance at zero magnetic field, plotted at the right axis. \textbf{d}, Magnetic field dependence of $R_{xy}$ at $\varphi$ = 45$\degree$. Open symbols show half of the PHE amplitude obtained by fitting the curves in \textbf{c} with a sin($\varphi$)cos($\varphi$) dependence. Dashed lines show fits to $n_{QP}/(1+exp(H_1/(H+H_0)))$.}
\label{Figure 2}
\end{figure}

Next, we examine the temperature and field dependence of PHE to better understand its origins and extent. In Figure~\ref{Figure 2}a, we plot the PHE signal as a function of temperature through the superconducting transition window (85 - 90~K) and at 100~K as a reference. PHE is virtually absent in the normal state (100~K) and appears only when YBCO transitions to the superconducting state, with the largest signal recorded at 87.5~K. Fig.~\ref{Figure 2}b shows the temperature dependence of PHE (left axis) plotted together with $R_{xx}$ (right axis). The onset of large PHE coincides with the onset of the superconducting transition, while the peak value occurs precisely at the end of the transition. Upon lowering the temperature further, the signal slowly decays and vanishes below 80~K, presumably due to the total suppression of transverse scattering events when Cooper pairs entirely dominate the conduction.

Fig.~\ref{Figure 2}c shows the magnetic field dependence of PHE at 88~K in the 0 - 8~T field range. PHE monotonously increases with the external field strength, evidencing that it is a field-driven effect rather than a magnetic order. By performing similar measurements at various temperatures within the superconducting transition (86 - 90~K), we obtain the field dependence of PHE shown in Fig.~\ref{Figure 2}d. The field dependence appears to follow a distinct power law that varies with temperature and tends to saturate at high fields. This behaviour can be modelled by fitting a phenomenological equation of the form $R_{xy} = n_{QP}/(1+exp(H_1/(H+H_0)))$ with the assumption that PHE arises due to spin-polarised quasiparticles, as discussed later. In this context, $n_{QP}$ represents the saturation density of spin-polarised quasiparticles, $H_1$ indicates the rate at which spin-polarised quasiparticle density increases as the magnetic field is applied, and $H_0$ denotes the critical field at which the quasiparticle density begins to rise significantly. This equation effectively captures the behaviour of superconducting quasiparticles in response to an applied magnetic field, modelling the smooth transition from a superconducting state (with negligible spin-polarised quasiparticles) to a normal state (with a high density of such quasiparticles) as the magnetic field increases.

\section*{Magnetoresistance and anomalous Hall effects}

\begin{figure*}[h]
\centering
\includegraphics[width=0.9\textwidth]{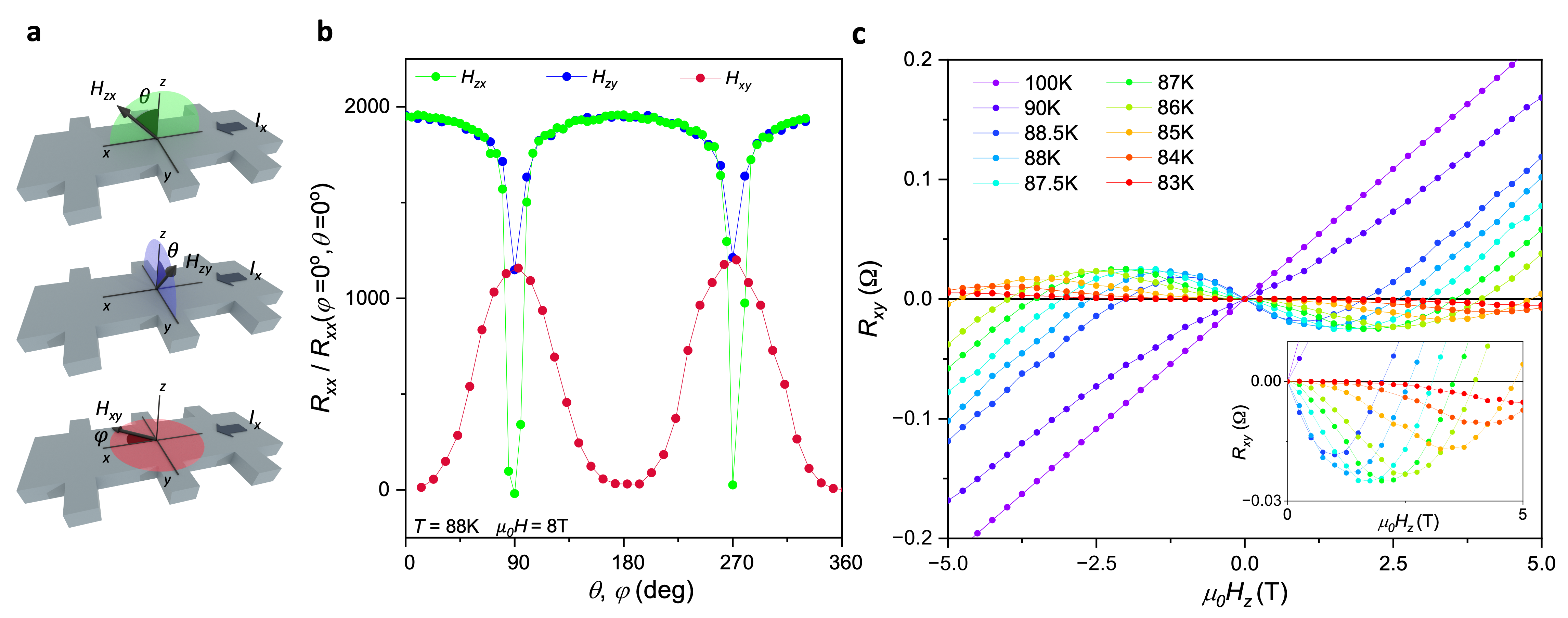}
\caption{\textbf{Out-of-plane longitudinal and Hall resistance.} \textbf{a,b}, Angular dependence of the longitudinal magnetoresistance, normalized to its value at $R_{xx}(\varphi =0,\theta =0)$, obtained at 88K and 8T by rotating the magnetic field along the different planes illustrated in \textbf{a}. \textbf{c}, Out-of-plane magnetic field dependence of odd component of the Hall magnetoresistance, $R_{xy}$, measured at different temperatures for a typical YBCO device of 50nm. The inset shows a zoom of the curves presented in the main panel.} 
\label{Figure 3}
\end{figure*}

To gather further signatures of the anomalous transport behaviour at the superconducting transition temperature, we measure the angular dependence of $R_{xx}$ in all three planes (i.e., \textit{xy}, \textit{zx} and \textit{zy} as schematically shown in Figure~\ref{Figure 3}a) at 88~K and 8~T. The results are shown in Fig.~\ref{Figure 3}b. The resistance is high when the magnetic field has a substantial out-of-plane component, i.e., far away from $\theta$ = 90\degree or 270\degree. When the field approaches 90\degree and 270\degree, $R_{xx}$ undergoes sharp drops to respective resistance values for the field aligned along \textit{x} and \textit{y}. Notably, $R_{xx}$ does not follow a geometrical relationship with the field (i.e., simple $\propto$ cos$^2$$\theta$ dependence) but transits to a low-resistance state when the in-plane field component is significantly larger than the out-of-plane one, satisfied when $\theta \approx 90\degree \pm$ 10\degree or 270\degree $\pm$ 10\degree. These measurements reveal that a strictly in-plane field is required for a large PHE, and the out-of-plane field is detrimental and drives the film into a high-resistance state with a weak response to the magnetic field. 

A further hint of the anomalous transport behaviour comes from the Hall effect measurements with an out-of-plane swept field shown in Fig.~\ref{Figure 3}c. In this analysis, we plot the odd component of $R_{xy}$ to eliminate the effects of vortex motion or contact misalignment. In the normal state, $T$= 100~K, $R_{xy}$ exhibits a strictly linear relationship with the applied magnetic field, described by $R_{xy}={H_{z}/nqt}$, where $q$ and $n$ are the normal state charge and carrier densities and $t$ is the film thickness. As the temperature approaches $T_c$, nonlinear behaviour emerges, leading to oscillations in $R_{xy}$, and a sign reversal of the Hall signal at low field is observed. In particular, a minimum signal appears, characterised by a dip in the $R_{xy}$ data (see the inset in Fig.~\ref{Figure 3}c) when the slope changes from negative (at low fields) to positive (in high fields). The nonlinearity in the $R_{xy}$ data can be associated with superconducting fluctuations and quasiparticle scattering, which are intricately connected to the specific characteristics of the Fermi surface and are highly dependent on carrier doping levels \cite{Zhao2019,Puica2004}. 

\begin{figure*}[h]
\centering
\includegraphics[width=0.9\textwidth]{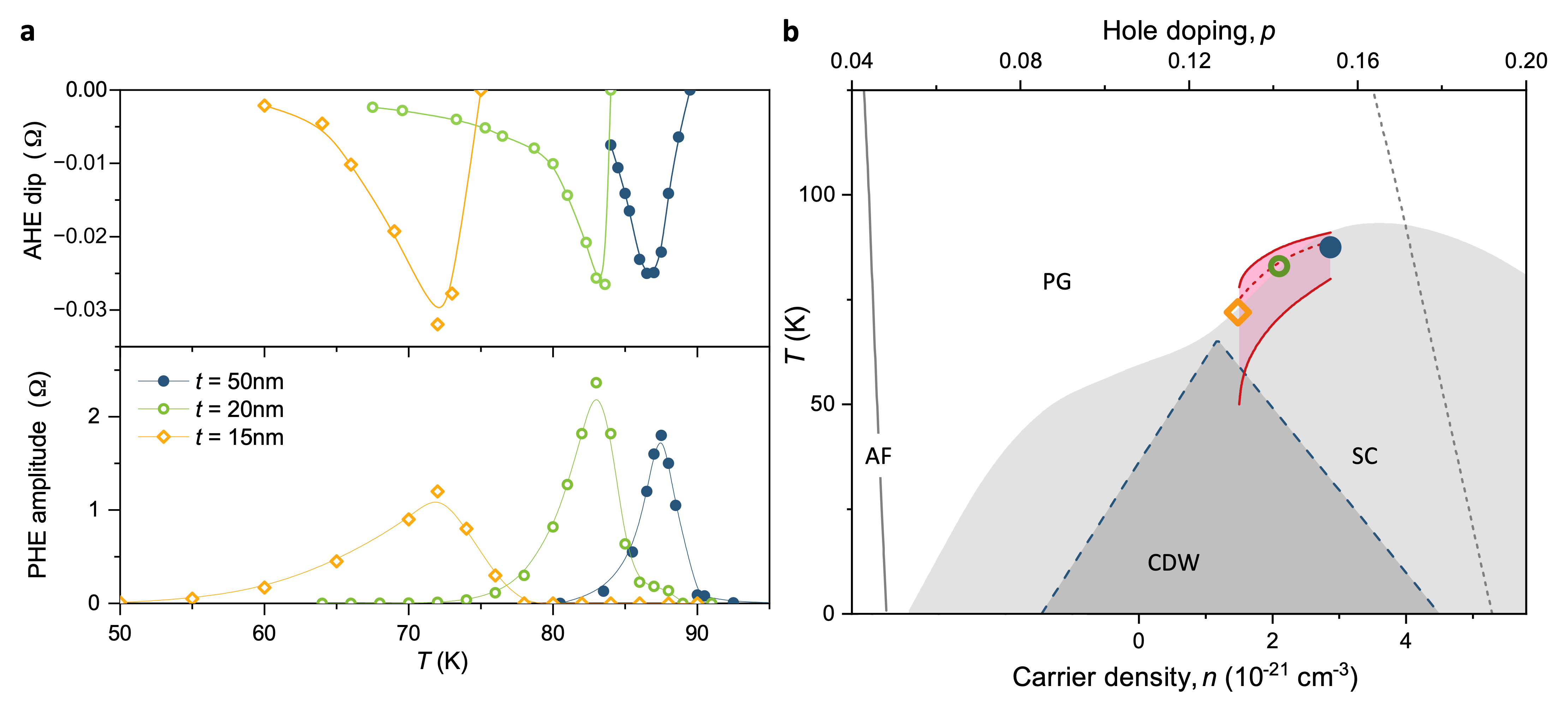}
\caption{\textbf{Doping dependent Hall measurements.} \textbf{a,b}, Temperature dependence of the AHE dip and PHE amplitude obtained for samples with different thicknesses: blue solid dots for 50nm, green open dots for 20nm, and yellow open diamonds for 15nm. \textbf{b} Temperature of the PHE amplitude peak (AHE dip) and carrier density for the three samples shown in \textbf{a}. The points have been plotted in the temperature-doping phase diagram of YBCO highlighting the different phases appearing: antiferromagnetic (AF) order, pseudogap (PG), superconductivity (SC) and charge density wave (CDW) order \cite{Badoux2016,Keimer2015}. The red dashed area shows the temperature range in which the PHE is observed.} 
\label{Figure 4}
\end{figure*}

Considering that the dips observed in Fig.~\ref{Figure 3}c might be related to the anomalous transport behaviour evidenced in the magnetoresistance and PHE, we quantify their values as a function of temperature and plot them in Figure~\ref{Figure 4}a - upper panel, together with the temperature evolution of PHE in the lower panel. Note that we now extended the analysis to lower thicknesses of the YBCO, namely 20 and 15 nm, which possess reduced oxygen doping leading to films with significantly lower superconducting transition temperatures \cite{Arpaia2017,Li2020}. The dips in AHE and PHE show a remarkable correlation in all three films, pinpointing their common origin.  Fig.~\ref{Figure 4}b shows the temperature versus carrier density phase diagram of YBCO \cite{Badoux2016,Keimer2015} with the shaded region the extent of PHE with the largest value plotted as data points. In all cases, the effect is most pronounced in the vicinity of $T_c$, and extends to a broader region as the doping level decreases. The localisation of the effect in a narrow range near $T_c$ provides strong evidence that the observed PHE is linked to quasiparticles. In this context, a coherence peak in conductivity appears near the critical temperature, which is attributed to the divergence of the quasiparticle density of states \cite{Gao1996}. Furthermore, the quasiparticle recombination lifetime in cuprate superconductors is significantly affected by the carrier doping, exhibiting a local maximum in a small temperature window near $T_c$ \cite{Hinton2016}. Additionally, spin-to-charge conversion effects mediated by quasiparticles are notably enhanced close to $T_c$ \cite{Jeon2020}.

\section*{Signatures of the spin-orbit coupling}

AMR and PHE are typically observed in magnetic materials and non-magnetic quantum materials with topological properties \cite{Li2022rev,Zhong2023} or in single crystals with broken rotational symmetry \cite{Wu2017}. The latter case can be easily tested by measuring devices patterned relative to different crystallographic orientations. The PHE in our films is independent of the device orientation, ruling out this effect. In the former case non-magnetic materials, the origin of the PHE is associated with a chiral anomaly in topological states owing to spin-textured bands \cite{Liao2020}. In particular, large AMR and PHE have been observed in topological superconductors exhibiting strong SOC \cite{Huang2018,Yang2021, Li2022,Feng2022}. In these systems, in contrast to what we observe, the effect persists at high temperatures well above the superconducting transition, and the in-plane anisotropic magnetoresistance is associated with topological surface states as an experimental signature of nontrivial Berry curvature and chiral anomaly. We propose that the PHE in our films is caused by quasiparticles excited in a narrow temperature window around $T_c$, which become spin-polarised due to the Zeeman splitting of majority and minority bands induced by an in-plane magnetic field, as postulated earlier \cite{Linder2015,Bergeret2018}. Field-induced dephasing of spin-polarised quasiparticles may then give rise to transverse and longitudinal magnetoresistive responses. Such a plausible scenario naturally suggests the presence of strong spin-orbit interactions and topologically-protected spin-momentum locking of quasiparticles in YBCO, not previously proposed or experimentally observed. Recent spin- and angle-resolved photoemission spectroscopy (SARPES) studies of bismuth-based cuprates on the spin characteristics of quasiparticles have revealed unexpected nontrivial spin textures, indicating that SOC may play a significant role in the physics of cuprate superconductors \cite{Gotlieb2018,Iwasawa2023,Luo2024}. We now show compelling evidence that a similar electronic structure might be at play in YBCO and have a drastic influence on transport properties around the superconducting transition temperature.

\begin{figure*}[h]
\centering
\includegraphics[width=0.9\textwidth]{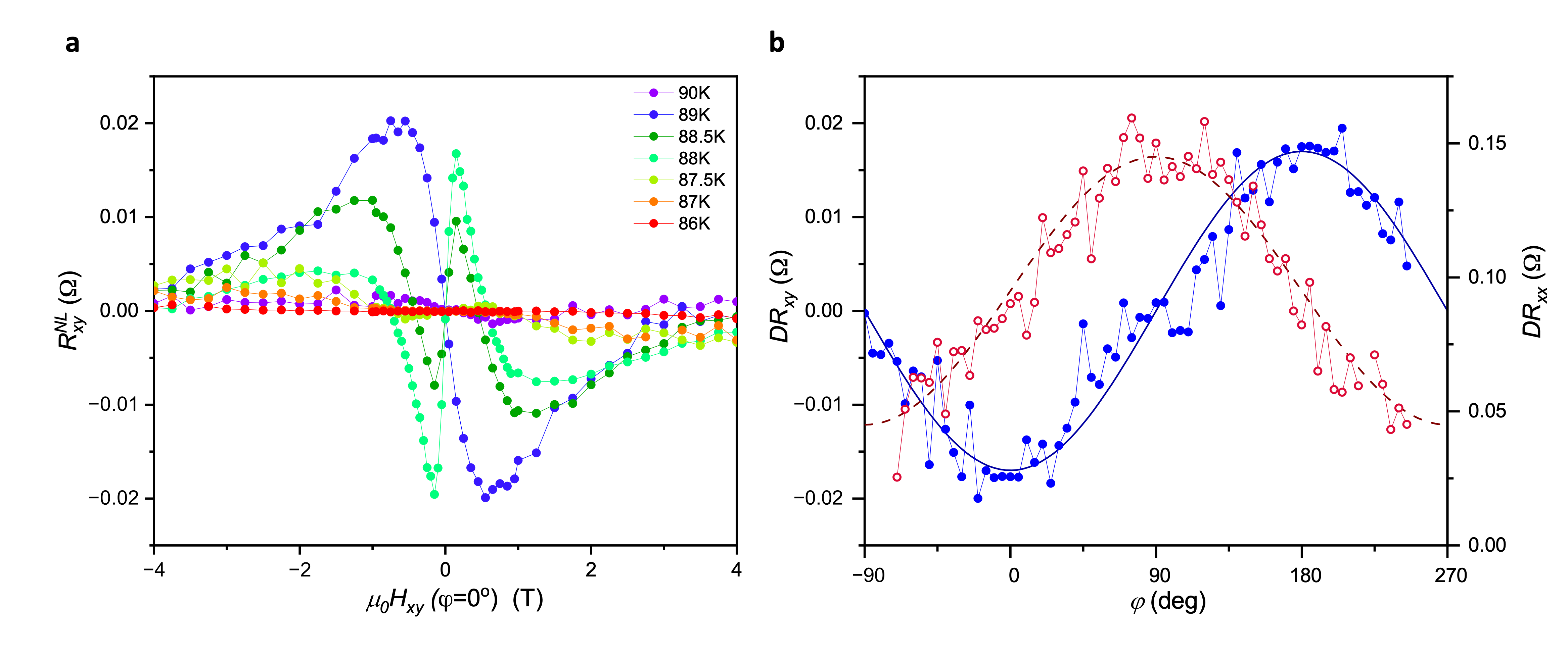}
\caption{\textbf{Nonlinear transverse and longitudinal resistances}. \textbf{a}, Magnetic field dependence of the non-linear PHE measured at $\varphi$ = 0$\degree$ and at various temperatures for a 50 nm YBCO device. \textbf{b} In-plane angular dependence of the non-linear Hall resistance (closed symbols, left axis) and magnetoresistance (open symbols, right axis) measured at 89K and 0.35T. Solid and dashed lines show the angular dependence of cos($\varphi$) and sin($\varphi$), respectively. } 
\label{Figure 5}
\end{figure*}

In systems with topologically protected spin-polarised transport, longitudinal and transverse resistances also entail nonlinear components \cite{He2018,He2019,Itahashi2020,Rao2021}. To investigate this aspect, we characterize the nonlinear response of YBCO by extending the analysis to the second order in the applied current, such that $V_{xy}$ =  $R_{xy}I_x$ + $R_{xy}^{NL}I_x^2$ and $V_{xx}$ =  $R_{xx}I_x$ + $R_{xx}^{NL}I_x^2$. $R_{xy}^{NL}$ and $R_{xx}^{NL}$ have then been determined by taking the difference of transverse and longitudinal resistances under positive, $I_x^+$, and negative, $I_x^-$, currents. Figure~\ref{Figure 5}a shows $R_{xy}^{NL}$ measured in the 86 - 90~K range for the 50 nm-thick YBCO as a function of the magnetic field applied parallel to the current ($\varphi=0\degree$). Clear $R_{xy}^{NL}$ signals emerge within the same temperature range where the large PHE was observed. Going down in temperature, $R_{xy}^{NL}$ first appears at 89K, increasing with the magnetic field up to 0.6~T and subsequently decreasing and disappearing at 4~T. Notably, a sign change of $R_{xy}^{NL}$ occurs when the temperature is decreased to 88~K, underlining competing effects. Fig.~\ref{Figure 5}b shows the angular dependence of the nonlinear transversal and longitudinal resistances obtained at 89~K, 0.35~T. $R_{xy}^{NL}$ and $R_{xx}^{NL}$ exhibit cos($\varphi$) and sin($\varphi$) dependences, respectively, with a periodicity of 2$\pi$, consistent with observations in non-magnetic topological insulators \cite{He2018,He2019,Rao2021}.

We attempt to qualitatively model $R_{xy}^{NL}$ and $R_{xx}^{NL}$ by considering spin-polarised quasiparticles interacting with an induced in-plane field mutually orthogonal to the current direction and the $z$-axis resulting from a bulk Rashba effect. This imprinted field could result from local structural fluctuations within the $\ce{CuO2}$ planes, creating a charge imbalance and breaking the bulk inversion symmetry \cite{Luo2024}. We postulate that the scattering rate of spin-polarised quasiparticles might be different when their polarisation is parallel and antiparallel to the intrinsic Rashba field, assumed to be along the $y$-axis. This would give rise to a bilinear (unidirectional) magnetoresistance depending on the current direction or field reversal and scaling linearly with the injected current \cite{He2018,vaz2020determining}. Analogous to the relationship between AMR and PHE, a similar effect is also expected in $R_{xy}^{NL}$ with a 90$\degree$ phase shift, as shown in Fig.~\ref{Figure 5}b. The initial increase of the signal is attributed to the field-induced generation of spin-polarised quasiparticles, which is necessary to generate the magnetoresistance asymmetry. Further field-driven decay underscores the interplay of external and intrinsic fields acting on quasiparticles. The sign change of the nonlinear signals could be linked to the modifications in the doping level and can be equated to changes in the binding energy; specifically, increased doping or a decrease in temperature yields a clockwise helicity, whereas decreased doping or an increase in temperature results in the opposite effect \cite{Luo2024}. This prediction aligns perfectly with the experimentally observed sign reversal of $R_{xy}^{NL}$. The simplistic model presented here, relying upon previous knowledge on topological materials and supported by extensive experimental data, captures the essential features of the nonlinear longitudinal and transverse resistance components. Further studies are required to unveil the exact mechanisms giving rise to specific signs and amplitudes of the effects and their field dependences.  

Linear and nonlinear magnetoresistive responses have gained renewed interest in solid-state physics as they contain invaluable information about the electronic structure at the Fermi level, SOC, spin-momentum locking, topology, and quantum metric in a wide variety of materials\cite{Sodeman2015,avci2015unidirectional,He2018,ma2019observation,Rao2021,gao2023quantum}. Extending these studies to superconductors possessing a richer palette of electronic states is essential to shed light on their transport properties. Therefore, our measurements provide a unique tool to investigate quasiparticles that are excited in various regions of the doping phase diagram where charge density wave and superconducting phases coexist (as illustrated in Fig. ~\ref{Figure 4}b). Specifically, quasiparticles excited from the superconducting phase are phase-coherent linear superpositions of normal-state electrons and holes, while charge-density wave quasiparticles consist of superpositions of electrons (or holes). Consequently, the recombination rate of these quasiparticles, as a function of temperature or magnetic field, shows a significant dependence on carrier doping \cite{Hinton2016}. This finding is particularly intriguing, as it suggests that adjusting doping levels may provide a mechanism for finely tuning the properties of YBCO and potentially the broader family of high-temperature superconductors, ultimately enabling functional quantum devices.
\section*{Summary and outlook}
High-temperature cuprate superconductors have long been viewed as materials characterised by strong electronic correlations but minimal SOC. However, recent observations of spin-momentum locking in bismuth-based cuprate superconductors suggest a more intricate spin landscape than previously recognised. Despite recent developments, the fundamental origins, intrinsic properties, and broader implications of this effect remain poorly understood. Our work provides direct experimental evidence of SOC in YBCO revealed through magnetoresistance and Hall effect measurements near the superconducting transition. These findings uncover a spin-dependent transport regime mediated by spin-polarised quasiparticles and challenge the conventional view of SOC’s insignificance in cuprates. The interplay between SOC, strong correlations, and superconductivity presents a rich frontier in condensed matter physics. Advancing experimental tools to investigate these interactions could lead to new insights into the physics of high-temperature superconductors. In particular, understanding how lattice symmetry and doping influence these interactions may hold the key to decoding the phase diagram of cuprates. Beyond fundamental interest, the emergence of spin-orbit-driven effects in high-temperature superconductors opens new opportunities for spintronic applications operating at liquid nitrogen temperatures.

\bibliography{References}

\begin{thebibliography}{10}
\expandafter\ifx\csname url\endcsname\relax
  \def\url#1{\burl{#1}}\fi
\expandafter\ifx\csname urlprefix\endcsname\relax\def\urlprefix{URL }\fi
\providecommand{\bibinfo}[2]{#2}
\providecommand{\eprint}[2][]{\url{#2}}
\providecommand{\doi}[1]{\url{https://doi.org/#1}}
\bibcommenthead

\bibitem{Ramirez_CMR1997}
\bibinfo{author}{Ramirez, A.~P.}
\newblock \bibinfo{title}{Colossal magnetoresistance}.
\newblock \emph{\bibinfo{journal}{Journal of Physics: Condensed Matter}} \textbf{\bibinfo{volume}{9}}, \bibinfo{pages}{8171} (\bibinfo{year}{1997}).

\bibitem{Phan_Magnetocaloric}
\bibinfo{author}{Phan, M.-H.} \& \bibinfo{author}{Yu, S.-C.}
\newblock \bibinfo{title}{Review of the magnetocaloric effect in manganite materials}.
\newblock \emph{\bibinfo{journal}{Journal of Magnetism and Magnetic Materials}} \textbf{\bibinfo{volume}{308}}, \bibinfo{pages}{325--340} (\bibinfo{year}{2007}).

\bibitem{Chakraborty2000}
\bibinfo{author}{Chakraborty, T.}
\newblock \bibinfo{title}{{Electron spin transitions in quantum Hall systems}}.
\newblock \emph{\bibinfo{journal}{Advances in physics}} \textbf{\bibinfo{volume}{49}}, \bibinfo{pages}{959--1014} (\bibinfo{year}{2000}).

\bibitem{Yan2017}
\bibinfo{author}{Yan, B.} \& \bibinfo{author}{Felser, C.}
\newblock \bibinfo{title}{ in \textit{{Topological Materials: Weyl Semimetals}}} (eds \bibinfo{editor}{Marchetti, M.~C.} \& \bibinfo{editor}{Sachdev, S.}) \emph{\bibinfo{booktitle}{Annual Review of Condensed Matter Physics}}, Vol.~\bibinfo{volume}{8} of \emph{\bibinfo{series}{Annual Review of Condensed Matter Physics}} \bibinfo{pages}{337--354} (\bibinfo{year}{2017}).

\bibitem{Qiu2018}
\bibinfo{author}{Qiu, Z.} \emph{et~al.}
\newblock \bibinfo{title}{{Spin colossal magnetoresistance in an antiferromagnetic insulator}}.
\newblock \emph{\bibinfo{journal}{Nature Materials}} \textbf{\bibinfo{volume}{17}}, \bibinfo{pages}{577} (\bibinfo{year}{2018}).

\bibitem{Lee2025}
\bibinfo{author}{Lee, T.} \emph{et~al.}
\newblock \bibinfo{title}{{Signatures of longitudinal spin pumping in a magnetic phase transition}}.
\newblock \emph{\bibinfo{journal}{Nature}} \textbf{\bibinfo{volume}{638}}, \bibinfo{pages}{106–111} (\bibinfo{year}{2025}).

\bibitem{YangRobinson2021}
\bibinfo{author}{Yang, G.}, \bibinfo{author}{Ciccarelli, C.} \& \bibinfo{author}{Robinson, J. W.~A.}
\newblock \bibinfo{title}{Boosting spintronics with superconductivity}.
\newblock \emph{\bibinfo{journal}{APL Materials}} \textbf{\bibinfo{volume}{9}}, \bibinfo{pages}{050703} (\bibinfo{year}{2021}).

\bibitem{Ohnishi2020}
\bibinfo{author}{Ohnishi, K.} \emph{et~al.}
\newblock \bibinfo{title}{Spin-transport in superconductors}.
\newblock \emph{\bibinfo{journal}{Applied Physics Letters}} \textbf{\bibinfo{volume}{116}} (\bibinfo{year}{2020}).

\bibitem{Kuzmanovic2020}
\bibinfo{author}{Kuzmanovic, M.}, \bibinfo{author}{Wu, B.~Y.}, \bibinfo{author}{Weideneder, M.}, \bibinfo{author}{Quay, C. H.~L.} \& \bibinfo{author}{Aprili, M.}
\newblock \bibinfo{title}{{Evidence for spin-dependent energy transport in a superconductor}}.
\newblock \emph{\bibinfo{journal}{Nature Communications}} \textbf{\bibinfo{volume}{11}}, \bibinfo{pages}{1--7} (\bibinfo{year}{2020}).

\bibitem{Quay2013}
\bibinfo{author}{Quay, C. H.~L.}, \bibinfo{author}{Chevallier, D.}, \bibinfo{author}{Bena, C.} \& \bibinfo{author}{Aprili, M.}
\newblock \bibinfo{title}{{Spin imbalance and spin-charge separation in a mesoscopic superconductor}}.
\newblock \emph{\bibinfo{journal}{Nature Physics}} \textbf{\bibinfo{volume}{9}}, \bibinfo{pages}{84--88} (\bibinfo{year}{2013}).

\bibitem{Hubler2012}
\bibinfo{author}{H{\"{u}}bler, F.}, \bibinfo{author}{Wolf, M.~J.}, \bibinfo{author}{Beckmann, D.} \& \bibinfo{author}{{V. L{\"{o}}hneysen}, H.}
\newblock \bibinfo{title}{{Long-range spin-polarized quasiparticle transport in mesoscopic al superconductors with a zeeman splitting}}.
\newblock \emph{\bibinfo{journal}{Physical Review Letters}} \textbf{\bibinfo{volume}{109}}, \bibinfo{pages}{1--5} (\bibinfo{year}{2012}).

\bibitem{Wakamura2015}
\bibinfo{author}{Wakamura, T.} \emph{et~al.}
\newblock \bibinfo{title}{{Quasiparticle-mediated spin Hall effect in a superconductor}}.
\newblock \emph{\bibinfo{journal}{Nature materials}} \textbf{\bibinfo{volume}{14}}, \bibinfo{pages}{675--679} (\bibinfo{year}{2015}).

\bibitem{Jeon2020}
\bibinfo{author}{Jeon, K.~R.} \emph{et~al.}
\newblock \bibinfo{title}{{Giant transition-state quasiparticle spin-Hall effect in an exchange-spin-split superconductor detected by nonlocal magnon spin transport}}.
\newblock \emph{\bibinfo{journal}{ACS Nano}} \textbf{\bibinfo{volume}{14}}, \bibinfo{pages}{15874--15883} (\bibinfo{year}{2020}).

\bibitem{Bergeret2018}
\bibinfo{author}{Bergeret, F.~S.}, \bibinfo{author}{Silaev, M.}, \bibinfo{author}{Virtanen, P.} \& \bibinfo{author}{Heikkil{\"{a}}, T.~T.}
\newblock \bibinfo{title}{Colloquium: Nonequilibrium effects in superconductors with a spin-splitting field}.
\newblock \emph{\bibinfo{journal}{Reviews of Modern Physics}} \textbf{\bibinfo{volume}{90}}, \bibinfo{pages}{1--23} (\bibinfo{year}{2018}).

\bibitem{Wolf2013}
\bibinfo{author}{Wolf, M.~J.}, \bibinfo{author}{H{\"{u}}bler, F.}, \bibinfo{author}{Kolenda, S.}, \bibinfo{author}{{V. L{\"{o}}hneysen}, H.} \& \bibinfo{author}{Beckmann, D.}
\newblock \bibinfo{title}{{Spin injection from a normal metal into a mesoscopic superconductor}}.
\newblock \emph{\bibinfo{journal}{Physical Review B - Condensed Matter and Materials Physics}} \textbf{\bibinfo{volume}{87}}, \bibinfo{pages}{1--7} (\bibinfo{year}{2013}).

\bibitem{Puica2004}
\bibinfo{author}{Puica, I.}, \bibinfo{author}{Lang, W.}, \bibinfo{author}{G{\"{o}}b, W.} \& \bibinfo{author}{Sobolewski, R.}
\newblock \bibinfo{title}{{Hall-effect anomaly near Tc and renormalized superconducting fluctuations in $\ce{YBaCuO_{7-x}}$}}.
\newblock \emph{\bibinfo{journal}{Physical Review B}} \textbf{\bibinfo{volume}{69}}, \bibinfo{pages}{1--7} (\bibinfo{year}{2004}).

\bibitem{Wordenweber2006}
\bibinfo{author}{W{\"{o}}rdenweber, R.}, \bibinfo{author}{Sankarraj, J.~S.}, \bibinfo{author}{Dymashevski, P.} \& \bibinfo{author}{Hollmann, E.}
\newblock \bibinfo{title}{{Anomalous Hall effect studied via guided vortex motion}}.
\newblock \emph{\bibinfo{journal}{Physica C: Superconductivity and its Applications}} \textbf{\bibinfo{volume}{434}}, \bibinfo{pages}{101--104} (\bibinfo{year}{2006}).

\bibitem{Wang1994}
\bibinfo{author}{Wang, Z.}, \bibinfo{author}{Dong, J.} \& \bibinfo{author}{Ting, C.}
\newblock \bibinfo{title}{{Unified Theory of Mixed State Hall Effect in Type-II Superconductors: Scaling Behavior and Sign Reversal}}.
\newblock \emph{\bibinfo{journal}{Physical Review Letters}} \textbf{\bibinfo{volume}{72}}, \bibinfo{pages}{3875--3878} (\bibinfo{year}{1994}).

\bibitem{Zhao2019}
\bibinfo{author}{Zhao, S.~Y.} \emph{et~al.}
\newblock \bibinfo{title}{{Sign-Reversing Hall Effect in Atomically Thin High-Temperature $\ce{Bi_{2.1}Sr_{1.9}CaCu_{2.0} O_{8+$\delta$}}$ Superconductors}}.
\newblock \emph{\bibinfo{journal}{Physical Review Letters}} \textbf{\bibinfo{volume}{122}}, \bibinfo{pages}{247001} (\bibinfo{year}{2019}).

\bibitem{Feigelman1995}
\bibinfo{author}{Feigelman, M.~V.}, \bibinfo{author}{Geshkenbeinn, V.~B.}, \bibinfo{author}{Larkin, A.~I.} \& \bibinfo{author}{Vinokur, V.~M.}
\newblock \bibinfo{title}{{Sign change of the flux-flow Hall effect in HTSC}}.
\newblock \emph{\bibinfo{journal}{JETP Letters}} \textbf{\bibinfo{volume}{62}}, \bibinfo{pages}{834--840} (\bibinfo{year}{1995}).

\bibitem{Wu2017}
\bibinfo{author}{Wu, J.}, \bibinfo{author}{Bollinger, A.~T.}, \bibinfo{author}{He, X.} \& \bibinfo{author}{Bo{\v{z}}ovi{\'{c}}, I.}
\newblock \bibinfo{title}{{Spontaneous breaking of rotational symmetry in copper oxide superconductors}}.
\newblock \emph{\bibinfo{journal}{Nature}} \textbf{\bibinfo{volume}{547}}, \bibinfo{pages}{432} (\bibinfo{year}{2017}).

\bibitem{Gotlieb2018}
\bibinfo{author}{Gotlieb, K.} \emph{et~al.}
\newblock \bibinfo{title}{{Revealing hidden spin-momentum locking in a high-temperature cuprate superconductor}}.
\newblock \emph{\bibinfo{journal}{Science}} \textbf{\bibinfo{volume}{362}}, \bibinfo{pages}{1271--1275} (\bibinfo{year}{2018}).

\bibitem{Iwasawa2023}
\bibinfo{author}{Iwasawa, H.} \emph{et~al.}
\newblock \bibinfo{title}{{Exploring spin-polarization in Bi-based high-Tc cuprates}}.
\newblock \emph{\bibinfo{journal}{Scientific Reports}} \textbf{\bibinfo{volume}{13}}, \bibinfo{pages}{13451} (\bibinfo{year}{2023}).

\bibitem{Luo2024}
\bibinfo{author}{Luo, H.} \emph{et~al.}
\newblock \bibinfo{title}{{Doping dependence of spin-momentum locking in bismuth-based high-temperature cuprate superconductors}}.
\newblock \emph{\bibinfo{journal}{Communications Materials}} \textbf{\bibinfo{volume}{5}}, \bibinfo{pages}{1--9} (\bibinfo{year}{2024}).

\bibitem{Raines2019}
\bibinfo{author}{Raines, Z.~M.}, \bibinfo{author}{Allocca, A.~A.} \& \bibinfo{author}{Galitski, V.~M.}
\newblock \bibinfo{title}{{Manifestations of spin-orbit coupling in a cuprate superconductor}}.
\newblock \emph{\bibinfo{journal}{Physical Review B}} \textbf{\bibinfo{volume}{100}}, \bibinfo{pages}{224512} (\bibinfo{year}{2019}).

\bibitem{Deutscher2014}
\bibinfo{author}{Deutscher, G.}
\newblock \bibinfo{title}{{Impact of pseudo-gap states on the pinning energy and irreversibility field of high temperature superconductors}}.
\newblock \emph{\bibinfo{journal}{APL Materials}} \textbf{\bibinfo{volume}{2}}, \bibinfo{pages}{096108} (\bibinfo{year}{2014}).

\bibitem{Arpaia2017}
\bibinfo{author}{Arpaia, R.} \emph{et~al.}
\newblock \bibinfo{title}{{Transport properties of ultrathin YBa$_2$Cu$_3$O$_{7-\delta}$ nanowires: a route to single photon detection}}.
\newblock \emph{\bibinfo{journal}{Physical Review B}} \textbf{\bibinfo{volume}{064525}}, \bibinfo{pages}{1--11} (\bibinfo{year}{2017}).

\bibitem{Li2020}
\bibinfo{author}{li, Z.} \emph{et~al.}
\newblock \bibinfo{title}{Suppression of superconductivity at the nanoscale in chemical solution derived $\ce{YBa_2Cu_3O_{7-d}}$ thin films with defective $\ce{Y_2Ba_4Cu_8O_{16}}$ intergrowths}.
\newblock \emph{\bibinfo{journal}{Nanoscale Advances}} \textbf{\bibinfo{volume}{2}}, \bibinfo{pages}{3384} (\bibinfo{year}{2020}).

\bibitem{Badoux2016}
\bibinfo{author}{Badoux, S.} \emph{et~al.}
\newblock \bibinfo{title}{{Change of carrier density at the pseudogap critical point of a cuprate superconductor}}.
\newblock \emph{\bibinfo{journal}{Nature}} \textbf{\bibinfo{volume}{531}}, \bibinfo{pages}{210--214} (\bibinfo{year}{2016}).

\bibitem{Keimer2015}
\bibinfo{author}{Keimer, B.}, \bibinfo{author}{Kivelson, S.~a.}, \bibinfo{author}{Norman, M.~R.}, \bibinfo{author}{Uchida, S.} \& \bibinfo{author}{Zaanen, J.}
\newblock \bibinfo{title}{{From quantum matter to high-temperature superconductivity in copper oxides.}}
\newblock \emph{\bibinfo{journal}{Nature}} \textbf{\bibinfo{volume}{518}}, \bibinfo{pages}{179--86} (\bibinfo{year}{2015}).

\bibitem{Gao1996}
\bibinfo{author}{Gao, F.} \emph{et~al.}
\newblock \bibinfo{title}{Quasiparticle damping and the coherence peak}.
\newblock \emph{\bibinfo{journal}{Physical Review B - Condensed Matter and Materials Physics}} \textbf{\bibinfo{volume}{54}}, \bibinfo{pages}{700--710} (\bibinfo{year}{1996}).

\bibitem{Hinton2016}
\bibinfo{author}{Hinton, J.~P.} \emph{et~al.}
\newblock \bibinfo{title}{{The rate of quasiparticle recombination probes the onset of coherence in cuprate superconductors}}.
\newblock \emph{\bibinfo{journal}{Scientific Reports}} \textbf{\bibinfo{volume}{6}}, \bibinfo{pages}{1--9} (\bibinfo{year}{2016}).

\bibitem{Li2022rev}
\bibinfo{author}{Li, T.}, \bibinfo{author}{Zhang, L.} \& \bibinfo{author}{Hong, X.}
\newblock \bibinfo{title}{{Anisotropic magnetoresistance and planar Hall effect in correlated and topological materials}}.
\newblock \emph{\bibinfo{journal}{Journal of Vacuum Science and Technology A}} \bibinfo{pages}{010807} (\bibinfo{year}{2022}).

\bibitem{Zhong2023}
\bibinfo{author}{Zhong, J.}, \bibinfo{author}{Zhuang, J.} \& \bibinfo{author}{Du, Y.}
\newblock \bibinfo{title}{{Recent progress on the planar Hall effect in quantum materials}}.
\newblock \emph{\bibinfo{journal}{Chin. Phys. B}} \textbf{\bibinfo{volume}{32}}, \bibinfo{pages}{047203} (\bibinfo{year}{2023}).

\bibitem{Liao2020}
\bibinfo{author}{Liao, Z.}, \bibinfo{author}{Jiang, P.}, \bibinfo{author}{Zhong, Z.} \& \bibinfo{author}{Li, R.~W.}
\newblock \bibinfo{title}{{Materials with strong spin-textured bands}}.
\newblock \emph{\bibinfo{journal}{npj Quantum Materials}} \textbf{\bibinfo{volume}{5}}, \bibinfo{pages}{1--10} (\bibinfo{year}{2020}).

\bibitem{Huang2018}
\bibinfo{author}{Huang, H.} \emph{et~al.}
\newblock \bibinfo{title}{{Giant anisotropic magnetoresistance and planar Hall effect in Sr0.06Bi2Se3}}.
\newblock \emph{\bibinfo{journal}{Applied Physics Letters}} \textbf{\bibinfo{volume}{22}}, \bibinfo{pages}{222601} (\bibinfo{year}{2018}).

\bibitem{Yang2021}
\bibinfo{author}{Yang, X.~C.} \emph{et~al.}
\newblock \bibinfo{title}{{Planar Hall effect in the quasi-one-dimensional topological superconductor}}.
\newblock \emph{\bibinfo{journal}{Physical Review B}} \textbf{\bibinfo{volume}{104}}, \bibinfo{pages}{1--9} (\bibinfo{year}{2021}).

\bibitem{Li2022}
\bibinfo{author}{Li, J.}, \bibinfo{author}{Wu, Z.} \& \bibinfo{author}{Feng, G.}
\newblock \bibinfo{title}{{Observation of Planar Hall Effect in a Strong Spin-Orbit Coupling Superconductor LaO0.5F0.5BiSe2}}.
\newblock \emph{\bibinfo{journal}{Journal of Superconductivity and Novel Magnetism}} \textbf{\bibinfo{volume}{35}}, \bibinfo{pages}{3521--3528} (\bibinfo{year}{2022}).

\bibitem{Feng2022}
\bibinfo{author}{Feng, G.}, \bibinfo{author}{Huang, H.}, \bibinfo{author}{Wu, Z.}, \bibinfo{author}{Han, Y.} \& \bibinfo{author}{Zhang, C.}
\newblock \bibinfo{title}{Planar hall effect and large anisotropic magnetoresistance in a topological superconductor candidate $\ce{Cu_{0.05}PdTe_2}$}.
\newblock \emph{\bibinfo{journal}{AIP Advances}} \textbf{\bibinfo{volume}{12}}, \bibinfo{pages}{1--6} (\bibinfo{year}{2022}).

\bibitem{Linder2015}
\bibinfo{author}{Linder, J.} \& \bibinfo{author}{Robinson, J.~W.}
\newblock \bibinfo{title}{{Superconducting spintronics}}.
\newblock \emph{\bibinfo{journal}{Nature Physics}} \textbf{\bibinfo{volume}{11}}, \bibinfo{pages}{307--315} (\bibinfo{year}{2015}).

\bibitem{He2018}
\bibinfo{author}{He, P.} \emph{et~al.}
\newblock \bibinfo{title}{{Bilinear magnetoelectric resistance as a probe of three-dimensional spin texture in topological surface states}}.
\newblock \emph{\bibinfo{journal}{Nature Physics}} \textbf{\bibinfo{volume}{14}}, \bibinfo{pages}{495--499} (\bibinfo{year}{2018}).

\bibitem{He2019}
\bibinfo{author}{He, P.} \emph{et~al.}
\newblock \bibinfo{title}{{Nonlinear Planar Hall Effect}}.
\newblock \emph{\bibinfo{journal}{Physical Review Letters}} \textbf{\bibinfo{volume}{123}}, \bibinfo{pages}{16801} (\bibinfo{year}{2019}).

\bibitem{Itahashi2020}
\bibinfo{author}{Itahashi, Y.~M.} \emph{et~al.}
\newblock \bibinfo{title}{Nonreciprocal transport in gate-induced polar superconductor $\ce{SrTiO_3}$}.
\newblock \emph{\bibinfo{journal}{Science Advances}} \textbf{\bibinfo{volume}{6}}, \bibinfo{pages}{eaay9120} (\bibinfo{year}{2020}).

\bibitem{Rao2021}
\bibinfo{author}{Rao, W.} \emph{et~al.}
\newblock \bibinfo{title}{{Theory for linear and nonlinear planar Hall effect in topological insulator thin films}}.
\newblock \emph{\bibinfo{journal}{Physical Review B}} \textbf{\bibinfo{volume}{103}}, \bibinfo{pages}{1--7} (\bibinfo{year}{2021}).

\bibitem{vaz2020determining}
\bibinfo{author}{Vaz, D.~C.} \emph{et~al.}
\newblock \bibinfo{title}{Determining the rashba parameter from the bilinear magnetoresistance response in a two-dimensional electron gas}.
\newblock \emph{\bibinfo{journal}{Physical Review Materials}} \textbf{\bibinfo{volume}{4}}, \bibinfo{pages}{071001} (\bibinfo{year}{2020}).

\bibitem{Sodeman2015}
\bibinfo{author}{Sodemann, I.} \& \bibinfo{author}{Fu, L.}
\newblock \bibinfo{title}{Quantum nonlinear hall effect induced by berry curvature dipole in time-reversal invariant materials}.
\newblock \emph{\bibinfo{journal}{Phys. Rev. Lett.}} \textbf{\bibinfo{volume}{115}}, \bibinfo{pages}{216806} (\bibinfo{year}{2015}).

\bibitem{avci2015unidirectional}
\bibinfo{author}{Avci, C.~O.} \emph{et~al.}
\newblock \bibinfo{title}{Unidirectional spin hall magnetoresistance in ferromagnet/normal metal bilayers}.
\newblock \emph{\bibinfo{journal}{Nature Physics}} \textbf{\bibinfo{volume}{11}}, \bibinfo{pages}{570--575} (\bibinfo{year}{2015}).

\bibitem{ma2019observation}
\bibinfo{author}{Ma, Q.} \emph{et~al.}
\newblock \bibinfo{title}{Observation of the nonlinear hall effect under time-reversal-symmetric conditions}.
\newblock \emph{\bibinfo{journal}{Nature}} \textbf{\bibinfo{volume}{565}}, \bibinfo{pages}{337--342} (\bibinfo{year}{2019}).

\bibitem{gao2023quantum}
\bibinfo{author}{Gao, A.} \emph{et~al.}
\newblock \bibinfo{title}{Quantum metric nonlinear hall effect in a topological antiferromagnetic heterostructure}.
\newblock \emph{\bibinfo{journal}{Science}} \textbf{\bibinfo{volume}{381}}, \bibinfo{pages}{181--186} (\bibinfo{year}{2023}).

\bibitem{Alcala2024}
\bibinfo{author}{Alcala, J.} \emph{et~al.}
\newblock \bibinfo{title}{{Tuning the superconducting performance of YBa2Cu3O7-$\delta$ films through field-induced oxygen doping}}.
\newblock \emph{\bibinfo{journal}{Scientific Reports}} \textbf{\bibinfo{volume}{14}}, \bibinfo{pages}{1939} (\bibinfo{year}{2024}).

\end{thebibliography}

\section*{Methods}

\textbf{Sample growth}\\
Epitaxial \ce{YBa_2Cu_3O_{7-x}} films of thicknesses 50~nm, 20~nm, and 15~nm were grown on top of \ce{LaAlO3} (LAO), \ce{SrTiO3} (STO) and \ce{(LaAlO3)_{0.3}-(Sr2AlTaO6)_{0.7}} (LSAT) single crystal substrates by pulsed laser deposition. Substrates were heated up to 810 \degree C, with an oxygen partial pressure of 0.3~mbar, with a target-to-substrate distance of 52.5 mm. Ablation of the target material was achieved by using a high fluence laser operating at 2~J/cm$^2$ and a frequency of 5 Hz. During cooling to room temperature, the samples were kept in an oxygen atmosphere of 1 Bar to ensure proper oxygenation of the deposited thin films. An under-doped 50 nm film was grown by post-annealing for 3 hours at 450 \degree C in a nitrogen atmosphere. Microstructural characterisation of YBCO thin films grown in similar conditions was reported elsewhere \cite{Alcala2024}.\\
\textbf{Device fabrication and measurements}\\
Four-terminal Hall devices were fabricated using standard photolithography and wet etching techniques with dimensions of 20-30 $\micro$m in length and 20-50 $\micro$m in width. Au contact electrodes with a thickness of 50 nm were deposited by sputtering. Transverse and longitudinal magnetoresistance measurements were conducted down to 10 K, using a Physical Property Measurement System (PPMS, Quantum Design) equipped with a sample rotator. The magnetic field was applied at various orientations by adjusting the sample's position relative to the field. A d.c. current of 5 mA was applied along the longitudinal direction. The critical temperature was determined from the maximum of the first derivative of $R_{xx}(T)$. The carrier density was calculated as $n= 1/(qR_H)$, where $R_H = t(\rm{d}R_{xy}/\rm{d}H_z)$ is the Hall coefficient $t$ the sample thickness and $\rm{d}R_{xy}/\rm{d}H_z$ the linear slope of the Hall resistance with the out-of-plane magnetic field, measured by applying a ramp of -5 to 5T. 

\section*{Acknowledgements}
We acknowledge financial support from the Spanish Ministry of Science and Innovation MCIN/ AEI /10.13039/501100011033/ through CHIST-ERA PCI2021-122028-2A co-financed by the European Union Next Generation EU/PRTR, the “Severo Ochoa” Programme for Centres of Excellence CEX2023-001263-S, HTSUPERFUN PID2021-124680OB-I00, co-financed by ERDF A way of making Europe. The Spanish Nanolito networking project (RED2022-134096-T). The European COST Action SUPERQUMAP (CA 21144) and the Scientific Services at ICMAB and the UAB PhD program in Materials Science. A.B. acknowledges support from MICIN Predoctoral Fellowship (PRE2019-09781). C.O.A acknowledges funding from the European Research Council (ERC) under the European Union’s Horizon 2020 research and innovation programme (project MAGNEPIC, Grant Agreement No. 949052) and the MAI-SKY (PID2021-125973OA-I00) project funded by MCIN/AEI/ 10.13039/501100011033/FEDER, UE. S.D. acknowledges funding from the Marie Sklodowska-Curie Actions (MSCA) under the European Union’s Horizon Europe research and innovation programme (project SPINDY, Grant Agreement No. 101106885). We thank Prof. Jason Robinson for fruitful discussions and valuable comments.

\newpage
\section*{Figures}

\end{document}